\documentclass[prd,twocolumn,superscriptaddress,altaffilletter,showpacs,nofootinbib]{revtex4}
\usepackage[dvips]{graphicx}
\usepackage{amsmath}

\newcommand{\be}{\begin{equation}}
\newcommand{\ee}{\end{equation}}
\newcommand{\bea}{\begin{eqnarray}}
\newcommand{\eea}{\end{eqnarray}}

\begin{document}

\title{QCD ghost dark energy cannot (even roughly) explain the main features of the accepted cosmological paradigm}

\author{Ricardo Garc\'{\i}a-Salcedo}\email{rigarcias@ipn.mx}\affiliation{Centro de Investigacion en Ciencia Aplicada y Tecnologia Avanzada - Legaria del IPN, M\'exico D.F., M\'exico.}

\author{Tame Gonzalez}\email{tamegc72@gmail.com}\affiliation{Departamento de Ingenier\'ia Civil, Divisi\'on de Ingenier\'ia, Universidad de Guanajuato, Gto., M\'exico.}

\author{Israel Quiros}\email{iquiros6403@gmail.com}\affiliation{Departamento de Matem\'aticas, Centro Universitario de Ciencias Ex\'actas e Ingenier\'ias, Universidad de Guadalajara, Guadalajara, Jal., M\'exico.}

\author{Michael Thompson-Montero}\email{mike_132001@hotmail.com}\affiliation{Departamento de F\'isica, Centro Universitario de Ciencias Ex\'actas e Ingenier\'ias, Universidad de Guadalajara, Guadalajara, Jal., M\'exico.}

\date{\today}

\begin{abstract}
We explore the whole phase space of the so called Veneziano/QCD ghost dark energy models where the dynamics of the inner trapping horizon is ignored and also the more realistic models where the time-dependence of the horizon is taken into consideration. We pay special attention to the choice of phase space variables leading to bounded and compact phase space so that no critical point of physical interest is missing. It is demonstrated that ghost dark energy is not a suitable candidate to explain the presently accepted cosmological paradigm since no critical point associated with matter dominance is found in the physical phase space of the model. A transient stage of matter dominance -- responsible for the observed amount of cosmic structure -- is an essential ingredient of the accepted cosmological paradigm. The above drawback is in addition to the well known problem with classical instability against small perturbations of the background density originated from negativity of the sound speed squared.
\end{abstract}

\pacs{05.45.-a, 05.70.Jk, 95.36.+x, 98.80.-k} 
\maketitle

\section{Introduction}

The nature of the present stage of accelerated expansion of the universe remains as one of the unexplained mysteries in physics. Assuming Einstein's general relativity (GR) is right leads to a multiplicity of models which explain the current inflationary stage of the cosmic evolution within appropriate data accuracy \cite{de-models,odintsov}. In spite of their relative success none of these models looks like a convincing explanation of the accelerated expansion due to several outcomes, among them: i) the cosmological constant problems, ii) the coincidence problem, iii) stability issues, etc. Ghost dark energy (GDE) models belong in this vast gallery \cite{urban,ariel,ohta,chinos,chinos',also,instability,alberto-rozas,other,gde-thermodynamics,iterms-also,global-behaviour}. Here the cosmological constant arises from the contribution of the so called Veneziano/QCD ghost fields \cite{low-e-qcd}. Although in flat Minkowski spacetime the QCD ghosts are unphysical and make no contribution, in curved/time-dependent backgrounds the cancellation of their contribution to the vacuum energy leave a small energy density $\rho\sim\Lambda^3_{QCD}H$, where $H$ is the Hubble parameter and $\Lambda_{QCD}\sim 100 MeV$ is the QCD mass scale \cite{urban,ariel,ohta,gde-motivation,gde-motivation'}. A very attractive feature of the QCD GDE models is linked with the fact that nothing behind the standard model (SM) of particles and GR is required to explain the origin of the dark energy. 

No matter how attractive the QCD GDE models could be, there is a serious objection against these cosmological models in connection with the stability issue. In Ref.\cite{chinos} the cosmological dynamics of a simple model where the GDE energy density is proportional to the Hubble parameter \cite{ohta} (here $8\pi G\equiv8\pi/m_{PL}^2\equiv c\equiv1$)

\bea \rho_{gde}=\alpha H,\;\;\alpha\sim\Lambda^3_{QCD},\label{sim-h}\eea was investigated. The authors found that the squared sound speed of GDE is negative in the model $c^2_s=dp_{gde}/d\rho_{gde}<0$, resulting in an instability against small perturbations of the background energy density $\delta\rho\propto e^{(\omega t)}\,e^{(i{\bf k}\cdot{\bf r})}$. The issue was investigated in detail in \cite{instability} where it was found that, due to non-positivity of the squared sound speed, both non-interacting and interacting GDE models are classically unstable against perturbations in flat and non-flat Friedmann-Robertson-Walker (FRW) backgrounds. Although several authors dismiss the instability problem raised by negativity of $c^2_s$ by arguing that the Veneziano ghost is not a physical propagating degree of freedom and the corresponding GDE model does not violate unitarity causality or gauge invariance \cite{alberto-rozas,ariel,global-behaviour}, others do not find the argument convincing. 

Other aspects of GDE models such as: i) equivalence with kinetic k-essence \cite{alberto-rozas}, ii) connection with Brans-Dicke, and $f(R)$ theories, and tachyons \cite{other}, and iii) thermodynamic issues \cite{gde-thermodynamics}, have been also investigated. In the latter case the QCD GDE energy density $\rho_{gde}$ is shown to be related with the radius of the trapping horizon $\tilde r_T$ \cite{gde-thermodynamics}: $$\rho_{gde}=\frac{\alpha(1-\epsilon)}{\tilde r_T}=\alpha(1-\epsilon)\sqrt{H^2+\frac{k}{a^2}},\;\epsilon\equiv\frac{\dot{\tilde r}_T}{2H\tilde r_T}.$$ If ignore the spatial curvature, as we do in this paper, the trapping horizon is coincident with the Hubble horizon $\tilde r_T=1/H$, and 

\be \rho_{gde}=\alpha(1-\epsilon)H,\;\;\epsilon=-\dot H/2H^2.\label{sim-h-e}\ee 

Usually the contribution from time-dependence of the radius of the horizon through $\epsilon$ is not considered \cite{urban,ariel,ohta,chinos,also,instability,alberto-rozas,other,gde-thermodynamics,iterms-also}. However, this is a particular case ($\epsilon=0$) which, strictly speaking, corresponds to dark energy dominance. Hence, this is no more than a convenient approximation which can not be intended to explain the whole extent of the cosmic history. In general, the contribution of $\epsilon$ to the dynamics has to be taken under consideration as it is done, for instance, in \cite{global-behaviour}. In that reference the global dynamical behavior of the universe accelerated by the QCD GDE, where $\rho_{gde}=\alpha(1-\epsilon)\sqrt{H^2+k/a^2}$, was investigated by using the dynamical systems tools. The authors considered the general situation when there is additional non-gravitational interaction between the QCD GDE and the cold dark matter (CDM) given by an interaction term -- source term in the continuity equations -- of the form $Q=3H(a\,\rho_{gde}+b\,\rho_{cdm})$, where the constants $a$, $b$, are adjustable free parameters.\footnote{This kind of linear interaction between the dark energy and the dark matter (and its particular cases) has been formerly studied within different contexts \cite{nm-coupling,i-term-pavon,ohta-nmc,pavon}.}

Depending on the values of these free parameters two critical points of physical relevance were found. Due to the choice of phase space variables in \cite{global-behaviour} ($\Omega_i\equiv\rho_i/3H^2$): $\mu=\Omega_{cdm}/\Omega_{gde}$, and $\epsilon$, both points $P_k:(\mu_k,\epsilon_k)$, $k=1,2$, are correlated with CDM/GDE-scaling behavior since $\mu_k\neq 0$, $$P_1:\left(\frac{a}{1-b},0\right),\;\;P_2:\left(-\frac{3a}{3b+1},1\right).$$ If remove the non-gravitational interaction ($a=b=0$) one would have, instead, $P_1:(0,0)$, and $P_2:(0,1)$, which correspond to GDE-dominated solutions ($\Omega_{gde}=1$). The additional non-gravitational interaction between GDE and CDM just transforms these critical points into GDE/CDM-scaling equilibrium points. In either case there are not critical points which can be associated with matter dominance. Hence, on the basis of this result the QCD GDE cosmological model may be ruled out as it is unable to explain the formation of structure in our universe. 

However, before making any conclusive argument against QCD GDE models on the basis of the results of the dynamical systems study in \cite{global-behaviour}, we should note that the study in that reference is unsatisfactory due to a problem with the choice of the variables of the phase space. Actually, if take into account (\ref{sim-h-e}) we can rewrite the variable $\mu$ as $\mu=\rho_{cdm}/\alpha(1-\epsilon)H$, i. e., $\mu$ and $\epsilon$ do not actually take independent values. As a consequence, for instance, the critical point $P_2$ may not exist since at $\epsilon=1$ the variable $\mu$ is undefined in general. Moreover, the variable $\mu$ above is unsatisfactory in yet another regard: it is unbounded $0\leq\mu<\infty$. This means, in particular, that critical points associated with CDM dominance -- if any -- are at infinity and may be lost. Perhaps for that reason the authors of the mentioned study apply, additionally, the qualitative technique of the so called 'nullcline' to complement the dynamical systems investigation. They found that, for a given region in parameter space and depending on the initial conditions, the end point of the cosmic evolution can be dominated by matter.

In the present paper we aim at a throughout investigation of the phase space dynamics of QCD GDE cosmological models. We consider separately the cases where the time-dependence of the radius of the horizon is not considered (\ref{sim-h}) and when it is taken into consideration (\ref{sim-h-e}). We pay special attention to a consistent choice of phase space variables leading, in particular, to bounded and compact phase space so that no critical point of physical relevance is missing. The results of our study will show that, as a matter of fact, since no critical point is found which may be correlated with transient matter dominance, the more realistic QCD GDE models where time-dependence of the horizon's radius is taken into account, can not explain the formation of structure in our universe. This drawback is in addition to the well known problem with classical instability against small perturbations of the background density originated from negativity of the sound speed squared \cite{chinos,instability}.

\section{Remarks on phase space analysis}

Usually the way to test the (theoretical/observational) viability of a given cosmological model is through using known solutions of the cosmological field equations or by seeking for new particular solutions that are physically plausible. However, in general, the cosmological field equations are very difficult to solve exactly and even when an analytic solution can be found it will not be unique but just one in a large set of them. This is not to talk about stability of given solutions. 

An alternative way around is to invoke the dynamical systems tools to extract useful information about the asymptotic properties of the model instead. In this regard knowledge of the critical (also equilibrium or fixed) points in the phase space corresponding to a given cosmological model is a very important information since, independent on the initial conditions chosen, the orbits of the corresponding autonomous system of ordinary differential equations (ODE) will always evolve for some time in the neighbourhood of these points. Besides, if the point were a global attractor, independent of the initial conditions, the orbits will always be attracted towards it either into the past or into the future. Going back to the original cosmological model, the existence of the critical points can be correlated with generic cosmological solutions that might really decide the fate and/or the origin of the cosmic evolution. 

The above interplay between a cosmological model and the corresponding phase space is possible due to an existing isomorphism between exact solutions of the cosmological field equations and points in the equivalent phase space spanned by given variables $(x,y,...)$. When we replace the original field variables $H$, $\rho_{cdm}$, $\rho_{gde}$, etc., by the phase space variables $$x=x(H,\rho_{cdm},...),\;y=y(H,\rho_{cdm},...),\;...,$$ we have to keep in mind that, at the same time, we trade the original set of non-linear second order differential equations in respect to the cosmological time $t$ (cosmological field equations), by a set of first order ordinary differential equations with respect to the variable $\tau=\ln a$: $$x'=f(x,y,...),\;y'=g(x,y,..),$$ etc. The most important feature of the latter autonomous system of ODE is that the functions $f(x,y,...)$, $g(x,y,...)$, $...$, do not depend explicitly on the parameter $\tau$. In other words, we are trading the study of the cosmological dynamics of $H=H(t)$, $\rho_{cdm}=\rho_{cdm}(t)$, $...$, by the study of the flux in $\tau$-parameter of the equivalent autonomous system of ODE. The critical points of this system $P_i:(x_i,y_i,...)$, i. e., the roots of the system of algebraic equations $$f(x,y,...)=0,\;g(x,y,...)=0,\;...,$$ correspond to solutions of the original system of cosmological equations. If consider small linear perturbations around $P_i$ $$x\rightarrow x_i+\delta x(\tau),\;y\rightarrow y_i+\delta x(\tau),\;...,$$ then these would obey the following system of coupled ODE:

\bea \begin{pmatrix} \delta x' \\ \delta y' \\ \vdots \end{pmatrix}=\begin{pmatrix} f_x & f_y & ...\\ g_x & g_y & ...\\ \vdots & \vdots & ...\end{pmatrix}_{P_i}\begin{pmatrix} \delta x \\ \delta y \\ \vdots \end{pmatrix},\label{01}\eea where the square matrix in the right-hand-side (RHS) of (\ref{01}) $J$ is the Jacobian (also linearization) matrix evaluated at $P_i$. If diagonalize $J$ then the coupled system of ODE (\ref{01}) gets decoupled:

\bea \begin{pmatrix} \delta\bar x' \\ \delta\bar y' \\ \vdots \end{pmatrix}=\begin{pmatrix} \lambda_1 & 0 & 0 &...\\ 0 & \lambda_2 & 0 & ...\\ \vdots & \vdots & \vdots & ...\\ 0 & 0 & ... & \lambda_n\end{pmatrix}\begin{pmatrix} \delta\bar x \\ \delta\bar y \\ \vdots \end{pmatrix},\label{02}\eea where $\lambda_1$, $\lambda_2$, etc., are the eigenvalues of the Jacobian matrix $J$, and the linear perturbations $\delta\bar x$, $\delta\bar y$, etc., are linear combinations of $\delta x$, $\delta y$, $...$: $\delta\bar x=c_{11}\delta x+c_{12}\delta y+...$, $\delta\bar y=c_{21}\delta x+c_{22}\delta y+...$, etc. Perturbations in Eq. (\ref{02}) are easily integrated:

\bea \delta\bar x(\tau)=\delta\bar x(0)\,e^{\lambda_1\tau},\;\delta\bar y(\tau)=\delta\bar y(0)\,e^{\lambda_2\tau},\;...\label{03}\eea In case the eigenvalues had non-vanishing imaginary parts the critical point $P_i$ is said to be spiral.\footnote{In general the eigenvalues can be complex numbers.} Depending on the signs of the real parts of the eigenvalues of $J$ the equilibrium point $P_i:(x_i,y_i,...)$ can be classified into:\footnote{In what follows we shall assume the point $P_i$ is an hyperbolic equilibrium point.} i) source point or past attractor if the real parts of all of the eigenvalues were positive quantities, ii) saddle point if at least one of the real parts of the eigenvalues were of a different sign (for example, $Re(\lambda_1)<0$, $Re(\lambda_2)>0$, etc.), and iii) future attractor if the real parts of all of the eigenvalues were negative quantities. In the last case the equilibrium point is stable against small perturbations $\delta x$, $\delta y$, etc., since these exponentially decay in $\tau$-time (see equations (\ref{03})).  

If a given equilibrium point $P_a:(x_a,y_a,...)$ were a global attractor, then, independent on the initial conditions chosen $x(\tau_0)=x_0,\;y(\tau_0)=y_0,...$, every orbit in the phase space will approach to $P_a$ into the future ($\{\tau: \tau>\tau_0\}$), i. e., the global (stable) attractor is the end point of any orbit in $\Psi$. On the contrary, if a given critical point $P_s:(x_s,y_s,...)$ were unstable, i. e., small perturbations around $P_s$ uncontrollably grow up with $\tau$, then this point were a past attractor or, also, the source point of any orbit in the phase space. For a third class of critical points, the so called ''saddle points'', depending on the initial conditions chosen, orbits in $\Psi$ can approach to this point, spend some time around it and then be repelled from it to finally approach to the stable attractor if it exists.\footnote{Our discussion here is oversimplified since, in general, critical points can be of many types, for instance, spiral, etc. Besides, there can be found also (un)stable manifolds such as cycles. To worsen things there can coexist several local attractors, saddle points, etc.} 

Suppose we have a typical phase portrait, composed of a source critical point $P_s$, a saddle point $P_*$, and a stable (global) attractor $P_a$. Each one of these points corresponds to a given solution of the original cosmological equations, $$H=H_s(z),\;H=H_*(z),\;H=H_a(z),$$ respectively. In the above expressions $z$ is the redshift which is related with $\tau$: $\tau=-\ln(z+1)$. A also typical orbit in the phase space will start at $P_s$ for $\tau=-\infty$, then will approach to $P_*$ and, after a finite (perhaps sufficiently long) $\Delta\tau$, will be repelled by $P_*$ to finally be attracted towards $P_a$. The parallel history in terms of the equivalent cosmological dynamics will be the following. The expansion starts with a Hubble parameter dynamics $H=H_s(z)$ then, as the Universe expands, the cosmic history enters a transient period characterized by the dynamics dictated by $H=H_*(z)$. After a perhaps long yet finite period $\Delta z$ the cosmic expansion will abandon the latter phase to enter into a stage which dynamics obeys $H=H_a(z)$ lasting for ever. 

We want to underline that in spite of the existing isomorphism between particular solutions of the cosmological field equations of a given model and points of the equivalent phase space, most of the exact particular solutions of the field equations can not be associated with critical points. These solutions can be picked up only under very specific initial conditions on the phase space orbits, i. e., these will be unstable solutions. If, for instance, there are not found fixed points which could be associated with CDM-dominated solution, this would mean that even if this is a particular exact solution of the cosmological field equations, it can not describe a matter-dominated stage of the cosmic expansion lasting for enough time as to produce the observed amount of cosmic structure. This result would dismiss a given model since it is not able to explain one of the main features of the accepted cosmological paradigm.

\section{Simplified GDE model where $\rho_{gde}\propto H$}\label{propto-h}

In this section we shall investigate the phase space dynamics of the simplified (approximate) GDE model when time-dependence of the Hubble horizon's radius is ignored \cite{gde-thermodynamics}. For the flat FRW universe filled with GDE, CDM, and radiation, the corresponding Friedmann equation is written as

\be 3H^2=\rho_{gde}+\rho_{cdm}+\rho_r,\label{friedman_eq}\ee where $\rho_{cdm}$ is the energy density of (pressureless) cold dark matter, $\rho_r$ is the energy density of radiation, and, $\rho_{gde}$ is the GDE energy density assumed to be given by Eq.(\ref{sim-h}). The energy conservation equations for the different components are:

\bea &&\dot\rho_{cdm}+3H\rho_{cdm}=0,\;\dot\rho_r+4H\rho_r=0,\notag \\
&&\dot\rho_{gde}+3H\rho_{gde}(1+\omega_{gde})=0, \label{conservation_law}\eea where $\omega_{gde}$ is the GDE EOS parameter. The definition for the GDE energy density (\ref{sim-h}), together with Eq.(\ref{conservation_law}), yield to the following relationship: 

\be \frac{\dot\rho_{gde}}{H\rho_{gde}}=\frac{\dot H}{H^2}=-3(1+\omega_{gde}).\label{relation}\ee 

It will be useful to have several quantities written in terms of the dimensionless parameter of GDE energy density $\Omega_{gde}\equiv\rho_{gde}/3H^2$ and of the dimensionless energy density parameter of the radiation component $\Omega_r\equiv\rho_r/3H^2$. In this regard, by taking the time derivative of the Friedmann equation (\ref{friedman_eq}) and, considering equations (\ref{friedman_eq}), and (\ref{conservation_law}), one is left with

\be 3(1+\omega_{gde})=\frac{3-3\Omega_{gde}+\Omega_r}{2-\Omega_{gde}}=-\frac{\dot H}{H^2}.\label{useful}\ee Hence, for the GDE state parameter one obtains $\omega_{gde}=-(3-\Omega_r)/[3(2-\Omega_{gde})]$, while for the deceleration parameter: $q=2+3\omega_{gde}$.

In order to put the cosmological equations in the form of an autonomous system of ODE, we choose appropriate phase space variables: $\Omega_{gde}$ and $\Omega_r$, or, for sake of simplicity and compactness of writing,

\be x\equiv\Omega_{gde},\;\;y\equiv\Omega_r.\label{xy}\ee After this choice the following autonomous system of ODE can be obtained:

\be x'=x\left[\frac{3-3x+y}{2-x}\right],\;y'=-2y\left[\frac{1+x-y}{2-x}\right],\label{ode}\ee where the tilde denotes derivative with respect to the variable $\tau\equiv\ln a$ ($d\tau=Hdt$). The phase space relevant to the present study is given by the following bounded triangular region in ($x,y$)-plane: 

\bea &&\Psi=\{(x,y):\,0\leq x\leq 1,\;0\leq y\leq 1,\nonumber\\
&&\;\;\;\;\;\;\;\;\;\;\;\;\;\;\;\;\;\;\;\;\;\;\;\;\;\;\;\;\;\;\;\;\;\;\;\;0\leq x+y\leq 1\}.\label{ph-sp}\eea In terms of the above variables, the Friedmann equation can be written in the form of the following constraint: $\Omega_{cdm}=1-x-y$, besides, the GDE EOS parameter can be written as $\omega_{gde}=(y-3)/[3(2-x)]$.

Three equilibrium points $P_{c_i}:(x_{c_i},y_{c_i})$ can be found in $\Psi$, which correspond to different phases of the cosmic evolution:

\begin{enumerate}

\item Radiation-dominated phase: \be P_r:(0,1),\;\Omega_{cdm}=0,\;\Omega_r=1,\;\Omega_{gde}=0.\nonumber\ee This is a decelerating expansion solution ($q=1$). The eigenvalues of the linearization matrix corresponding to this equilibrium point are: $\lambda_1=2$, $\lambda_2=1$, so that it is a unstable critical point (past attractor) in $\Psi$. The GDE state parameter is $\omega_{gde}=-1/3$.
  
\item CDM-dominated phase: \be P_m:(0,0),\;\Omega_{cdm}=1,\;\Omega_r=0,\;\Omega_{gde}=0.\nonumber\ee This phase of the cosmic evolution is characterized also by decelerated expansion ($q=1/2$). The existence of this solution is necessary for the formation of the observed amount of structure. The eigenvalues of the corresponding linearization matrix are: $\lambda_1=-1$, $\lambda_2=3/2$, so that it is a saddle critical point in $\Psi$. For the GDE state parameter we obtain $\omega_{gde}=-1/2$.
  
\item GDE-dominated, de Sitter phase: \be P_{dS}:(1,0),\;\Omega_{cdm}=0,\;\Omega_r=0,\;\Omega_{gde}=1.\nonumber\ee This late-time phase corresponds to an inflationary solution ($q=-1$). The eigenvalues of the linearization matrix for $P_{dS}$ are: $\lambda_1=-4$, $\lambda_2=-3$, so that the solution is a future attractor. This equilibrium point mimics cosmological constant behavior since $\omega_{gde}=-1$.

\end{enumerate}
  
The global structure of the phase space shows that orbits in $\Psi$ converge into the $\tau$-past towards the radiation-dominated stage. In the forward $\tau$-direction, depending on the initial conditions, these orbits evolve for some time in the neighborhood of the saddle matter-dominated point, until they are repelled from this point to, finally, converge towards the GDE-dominated future de Sitter attractor. This model correctly describes the fundamental stages of the cosmic evolution arising within the presently accepted cosmological paradigm: i) a stage of radiation domination from which, ii) a period of matter-radiation equality and subsequent matter domination emerge, followed by iii) a late-time inflationary stage.

In spite of the convenient structure of the phase space, the present model of QCD GDE where $\rho_{gde}=\alpha H$, i. e., where $\epsilon=0$ and time-dependence of the horizon is ignored, is just an approximation which is valid only whenever the ghost dark energy dominates the cosmic evolution. Hence one should not expect that this is a good model to explain the whole cosmic history. This means that we should not take too much seriously critical points other than the GDE-dominated one. As a matter of fact, in the next section where the more physically involved model with time-dependent horizon is explored, $\rho_{gde}=\alpha H(1-\epsilon)$, we shall see that only two critical points are found: i) the one related with ghost dark energy dominance, and ii) the other associated with an empty and static universe.

\section{Time-dependent horizon}\label{time-d}

Within the context of QCD GDE models of accelerated expansion the contribution from time-dependence of the radius of the horizon through the quantity $\epsilon$ is usually ignored \cite{urban,ariel,ohta,chinos,also,instability,alberto-rozas,other,gde-thermodynamics,iterms-also}, however this assumption is no more than a convenient simplification of the model. In general, the contribution coming from the dynamics of the horizon is to be taken under consideration. In the present section we shall be considering how the time-dependence of the horizon impacts the asymptotic properties of the Veneziano ghost dark energy model. 

For sake of simplicity and in order to keep the phase space 2-dimensional, here we shall omit the radiation component in the cosmological equations:

\bea &&3H^2=\rho_{cdm}+\rho_{gde},\nonumber\\
&&\dot H=-\frac{1}{2}\rho_{cdm}-\frac{1}{2}(1+\omega_{gde})\rho_{gde}.\label{cosmo-eq}\eea

In the present case the energy density of the Veneziano GDE is given by Eq.(\ref{sim-h-e}) and the dynamics of the horizon is encoded in the quantity $\epsilon=-\dot H/2H^2$. The continuity equations (\ref{conservation_law}) hold true, however, in place of (\ref{relation}) one now gets 

\be \frac{\dot\rho_{gde}}{H\rho_{gde}}=\frac{\rho'_{gde}}{\rho_{gde}}=-3(1+\omega_{gde})=-\frac{\epsilon'}{1-\epsilon}-2\epsilon.\label{relation-1}\ee Besides, by taking the derivative of the Friedmann equation in (\ref{cosmo-eq}), and considering the definition of $\epsilon$, one gets $$4\epsilon=3\omega_{gde}\Omega_{gde}+3\;\Rightarrow\;\omega_{gde}=\frac{4\epsilon-3}{3\Omega_{gde}}.$$

In order to investigate the phase space dynamics of this model it is convenient to introduce the following bounded phase space variables: 

\bea \xi=\frac{3H}{3H+\alpha},\;\zeta=\frac{1}{2-\epsilon}=\frac{2H^2}{4H^2+\dot H}\,,\label{ps-var}\eea $0\leq\xi\leq 1$, $0\leq\zeta\leq 1$ (see below). In terms of these variables one has

\bea &&\Omega_{gde}=\frac{\alpha(1-\epsilon)}{3H}=\frac{(1-\xi)(1-\zeta)}{\xi\zeta},\nonumber\\
&&\Omega_{cdm}=1-\Omega_{gde}=\frac{\xi+\zeta-1}{\xi\zeta},\label{fried-const}\eea while 

\bea q=-1+2\epsilon=\frac{3\zeta-2}{\zeta},\;\omega_{gde}=\frac{\xi(5\zeta-4)}{3(1-\xi)(1-\zeta)}.\label{q,w}\eea The cosmological equations (\ref{cosmo-eq}) can be written as

\bea \dot H=-\frac{3}{2}H^2(1+\omega_{gde}\Omega_{gde})=-2H^2\left(\frac{2\zeta-1}{\zeta}\right),\label{useful-1}\eea where we have taken into account that $$\omega_{gde}\Omega_{gde}=\frac{5\zeta-4}{3\zeta}.$$ Equations (\ref{fried-const}), (\ref{q,w}) and (\ref{useful-1}) will be useful for the discussion below.

The following autonomous system of ODE is obtained out of (\ref{cosmo-eq}), (\ref{conservation_law}):

\bea &&\xi'=\frac{2\xi(1-\xi)(1-2\zeta)}{\zeta},\nonumber\\
&&\zeta'=\frac{(1-\xi-\zeta)(2-\zeta)+2\xi\zeta(2\zeta-1)}{1-\xi}.\label{ode-e}\eea In what follows we shall investigate in all detail the asymptotic structure of this model.

\subsection{Physical phase space}

\begin{figure}[t!]
\includegraphics[width=8cm]{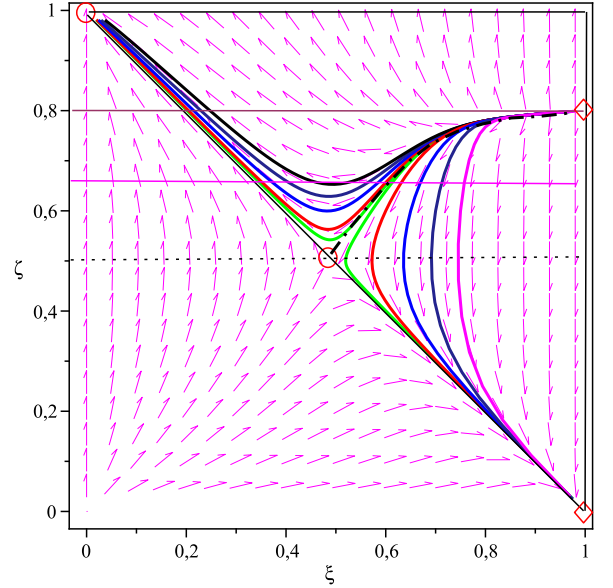}
\caption{Phase portrait for the autonomous system of ODE (\ref{ode-e}). The physical phase space $\Psi_\epsilon$ is the triangular region with vertexes at $(0,1)$, $(1,0)$ and $(1,1)$. Small circles at i) the geometrical center of the figure $P_{dS}:(\xi,\zeta)=(0.5,0.5)$, and at ii) the left-upper vertex $P_{E}:(0,1)$, enclose the two critical points of the autonomous system of ODE (\ref{ode-e}). Meanwhile the diamond-shaped forms enclose the non-critical (yet remarkable) points iii) $P_{MD}:(\xi,\zeta)=(1,0.8)$ which is associated with matter-dominated solution, and iv) $P_{BR}:(\xi,\zeta)=(1,0)$ corresponding to big rip event(s). Along the separatrix (dot-dashed curve joining the points $P_{dS}$ and $P_{MD}$): $\zeta=(7\xi-3)/(8\xi-3)$, the GDE equation of state parameter $\omega_{gde}=-1$. The doted horizontal line at $\zeta=0.5$ represents the so called 'phantom divide' line $\omega_{gde}=-1$. Points associated with accelerated expansion lie within that part of $\Psi_\epsilon$ below the horizontal line at $\zeta=0.66$, while those with negative $\omega_{gde}<0$ are located below $\zeta=0.8$.}\label{fig1}
\end{figure}

Since we consider expansion only ($H\geq 0$), then $\xi\geq 0$. Besides, as long as $\rho_{gde}=\alpha H(1-\epsilon)\geq 0$ is non-negative, then $\epsilon\leq 1$ $\Rightarrow\;\zeta\leq 1$. In general $-\infty\leq\epsilon\leq 1$,\footnote{Notice that $\epsilon$ can be negative ($\dot H>0$) and is unbounded from below.} i. e., $0\leq\zeta\leq 1$. These constraints together with the constraint $0\leq\Omega_{gde}\leq 1$ lead to $0\leq\xi\leq 1$. The resulting physical phase space where to look for critical points of the autonomous system of ODE (\ref{ode-e}) is defined as the following compact and bounded triangular region in $\xi\zeta$-plane (see Fig.\ref{fig1}):

\bea \Psi_\epsilon=\{(\xi,\zeta):\,0\leq\xi\leq 1,\;0\leq\zeta\leq 1,\;\xi+\zeta\geq 1\},\label{ph-sp-e}\eea where it has been considered also the constraint $0\leq\Omega_{cdm}\leq 1$. The three edges of the triangle in $\Psi_\epsilon$ are given by: 

\begin{enumerate}

\item The oblique edge 

\bea &&{\bf w}=\{(\xi,\zeta):\,\xi+\zeta=1,\,0<\xi< 1,\,0<\zeta<1\}\nonumber\\
&&\Rightarrow\;\Omega_{gde}=1,\;q=\frac{3\zeta-2}{\zeta},\;\omega_{gde}=\frac{5\zeta-4}{3\zeta},\label{w-edge}\eea i. e., $0<\zeta<1$ $\Rightarrow$ $-\infty<q<1$, $-\infty<\omega_{gde}<1/3$. All points in ${\bf w}$ are associated with GDE-dominated solutions. 

\item The (upper) horizontal edge 

\bea &&{\bf h}=\{(\xi,\zeta):\,0<\xi<1,\,\zeta=1\}\nonumber\\
&&\Rightarrow\;\Omega_{cdm}=1,\;\Omega_{gde}=0,\;q=1,\label{h-edge}\eea where $\omega_{gde}$ is undefined, however $$\omega_{gde}\Omega_{gde}=\frac{1}{3}\;\Rightarrow\;p_{gde}=H^2=\frac{\rho_{cdm}}{3}.$$ Hence, since according to (\ref{useful-1}) $\dot H=-2H^2$, while $\Omega_{cdm}=1$, then at points in ${\bf h}$ the ghost dark energy behaves as 'pure pressure'.

\item The (right-hand) vertical edge 

\bea &&{\bf v}=\{(\xi,\zeta):\,\xi=1,\,0<\zeta\leq 1\}\nonumber\\
&&\Rightarrow\;\Omega_{cdm}=1,\;q=\frac{3\zeta-2}{\zeta},\label{v-edge}\eea i. e., $0<\zeta\leq 1$ $\Rightarrow$ $-\infty<q\leq 1$, and $\omega_{gde}$ is undefined. As in the former case for points in ${\bf v}$, $\Omega_{gde}=0$ (no GDE energy density) while $$\omega_{gde}\Omega_{gde}=\frac{p_{gde}}{3H^2}=\frac{5\zeta-4}{3\zeta},$$ so that the GDE behaves as 'pure pressure' with $$p_{gde}=\frac{5\zeta-4}{3\zeta}\rho_{cdm},$$ but for the remarkable point $P_{m-d}:(\xi,\zeta)=(1,4/5)$, which is a standard matter-dominated (non-equilibrium) point.

\end{enumerate}

Points in the phase space corresponding to accelerated pace of the cosmic expansion lie below $\zeta=2/3$, within the triangular region (see Fig.\ref{fig1}):

\bea &&\Psi_{a-exp}(\subset\Psi_\epsilon)=\{(\xi,\zeta):\,1/3<\xi\leq 1,\nonumber\\
&&\;\;\;\;\;\;\;\;\;\;\;\;\;\;\;\;\;\;\;\;\;\;\;\;\;\;\;\;\;\;\;0\leq\zeta<2/3,\;\xi+\zeta\geq 1\},\label{acce-points}\eea while those with negative GDE EOS parameter $\omega_{gde}<0$ are located below the line $\zeta=4/5$.

Along the separatrix (dot-dashed curve in the figure \ref{fig1})

\bea \sigma=\left\{(\xi,\zeta):\,\zeta=\frac{7\xi-3}{8\xi-3},\;\frac{1}{2}<\xi<1\right\},\label{separatrix}\eea $$\omega_{gde}=-1,\;\Omega_{gde}=\frac{1-\xi}{7\xi-3},$$ i. e., $\sigma$ joints the matter-dominated point $(1,4/5)$ $\Rightarrow\;\Omega_{gde}=0$, with the GDE-dominated one at $(1/2,1/2)$ $\Rightarrow\;\Omega_{gde}=1$ (see below).


\begin{figure}[t!]
\includegraphics[width=4cm]{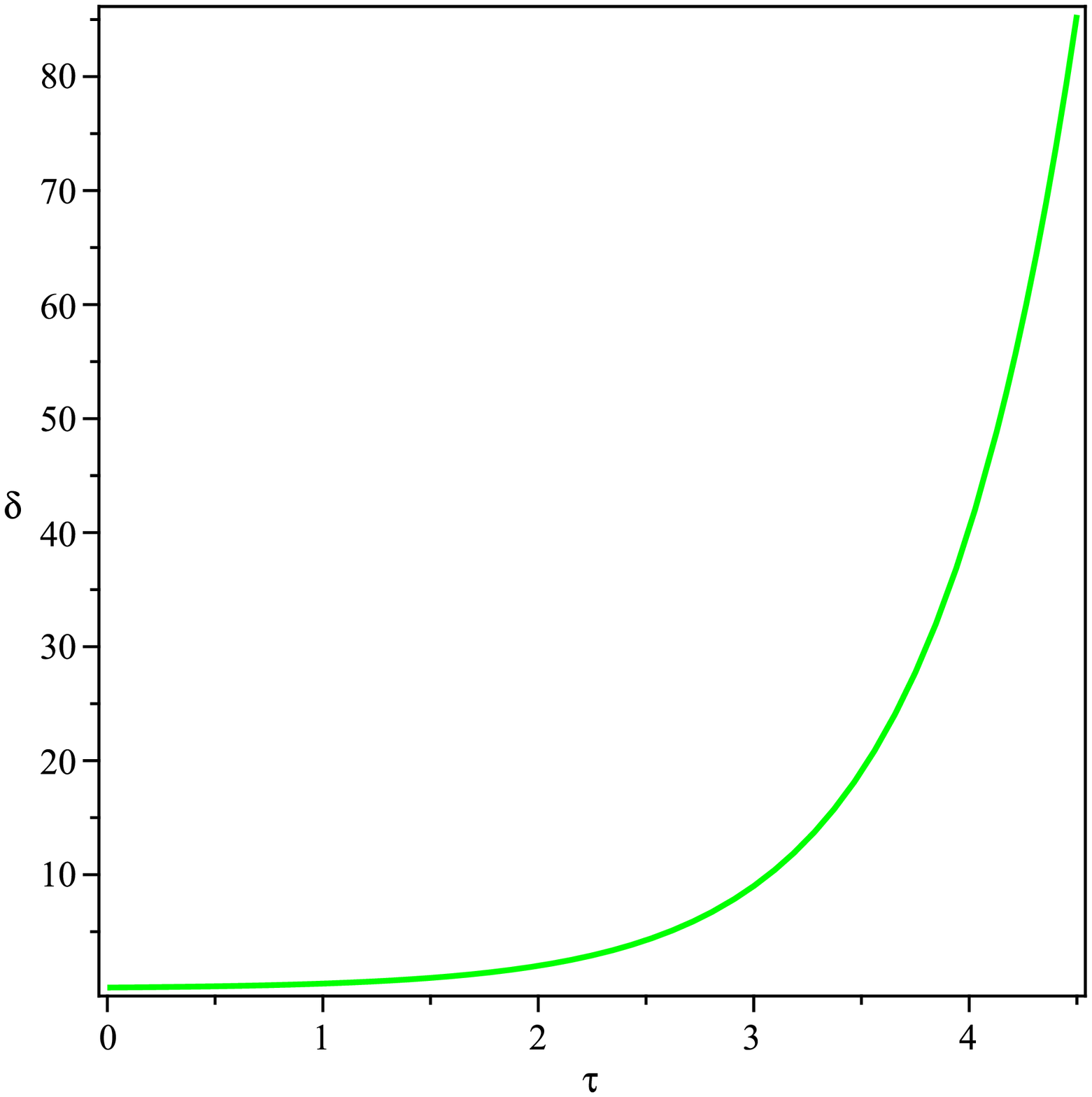}
\includegraphics[width=4cm]{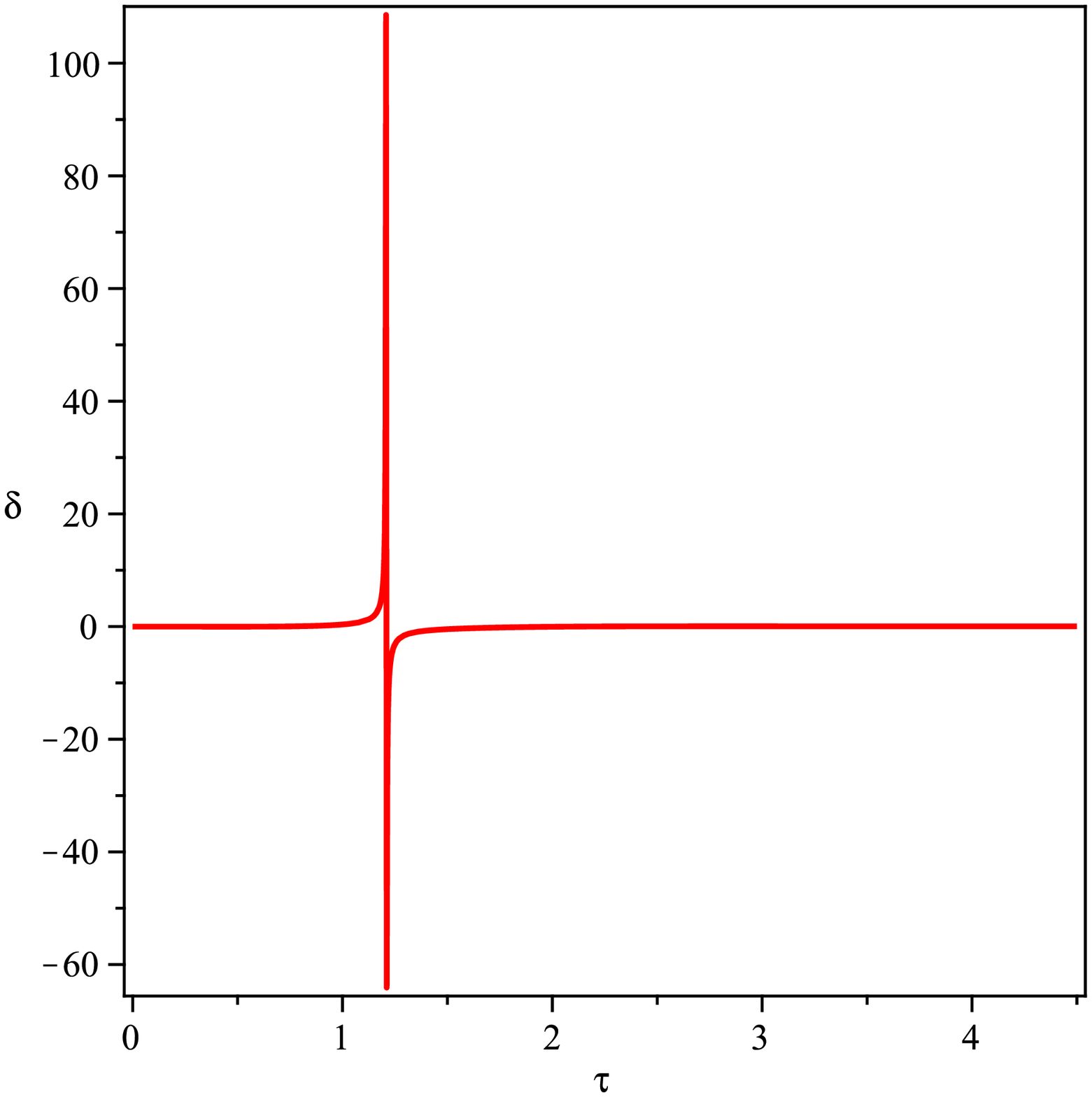}
\includegraphics[width=7cm]{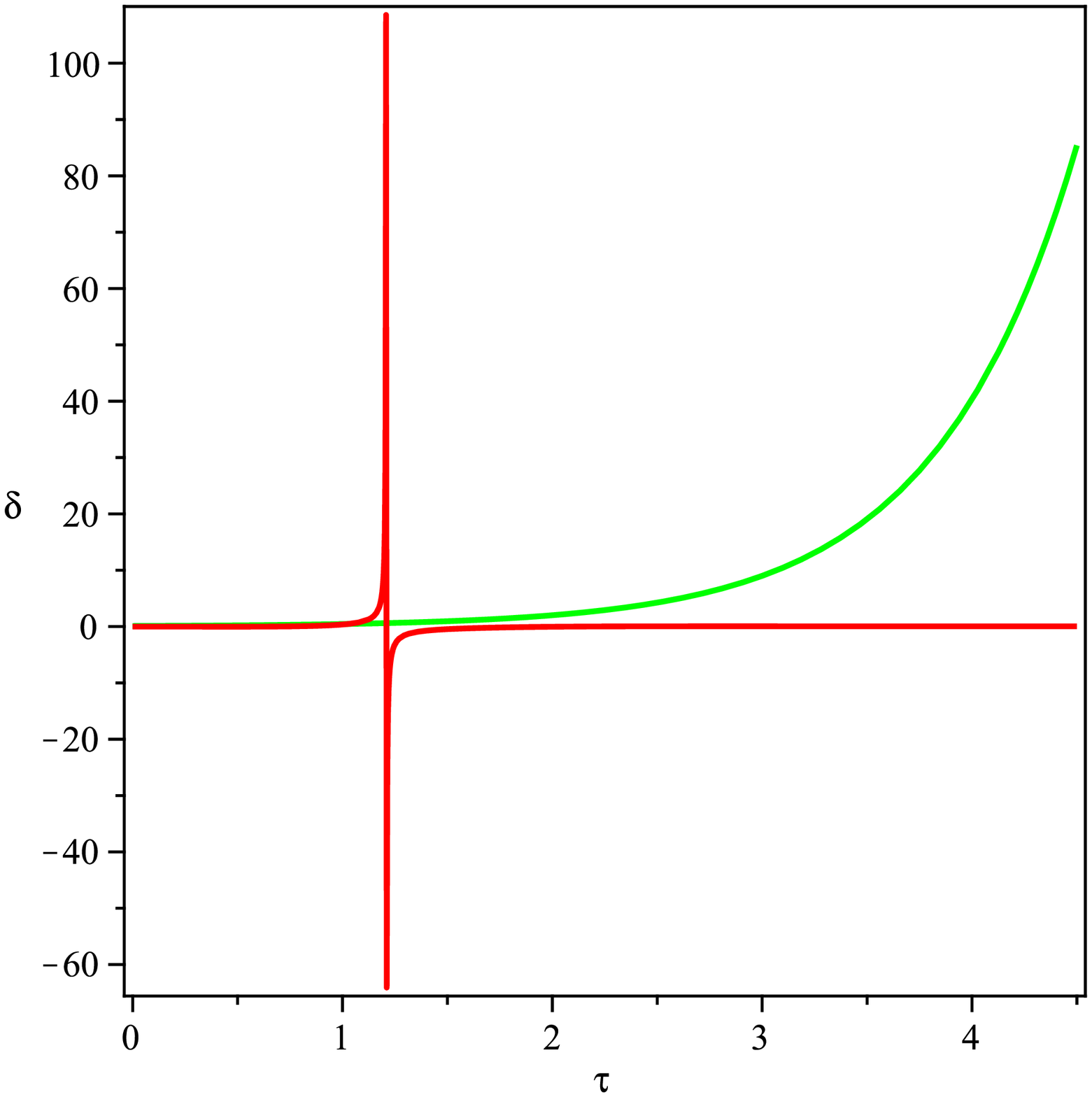}
\caption{Small perturbations in the neighborhood of the non-equilibrium point $P_{MD}:(\xi,\zeta)=(1,0.8)$ corresponding to the matter-dominated solution. In the upper left-hand panel the evolution of the small $\delta\xi$-perturbation in $\tau$-time is shown, while in the right-hand panel $\delta\zeta$ vs $\tau$ is depicted. We have chosen the following values of the constant parameters $\delta\xi(0)=0.1$, $C=0.1$ in equations (\ref{xi-pert}), (\ref{zeta-pert}). In the lower panel the evolution of both $\delta\xi(\tau)$ and $\delta\zeta(\tau)$ is simultaneously shown.}\label{fig2}
\end{figure}

\begin{figure}[t!]
\includegraphics[width=7cm]{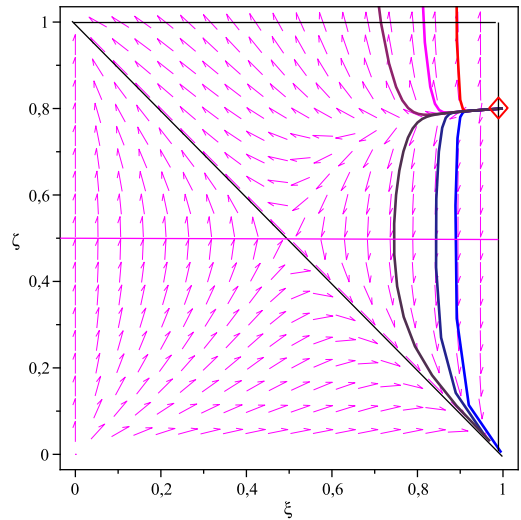}
\caption{Orbits generated by initial data in the neighborhood of the matter-dominated non-equilibrium point $P_{MD}:(1,0.8)$ very quickly approach either to pints in ${\bf h}$ which are not compatible with general relativity, or very early do the crossing of the phantom divide at $\zeta=0.5$ and approach to the big rip event. In either case none of these orbits approach to the saddle de Sitter equilibrium point.}\label{fig3}
\end{figure}


\subsection{Critical points in $\Psi_\epsilon$}

The equilibrium points $P_i:(\xi_i,\zeta_i)$ of the autonomous system of ODE (\ref{ode-e}) in the physical phase space $\Psi_\epsilon$ are: 

\begin{enumerate}

\item GDE-dominated de Sitter equilibrium point $P_{dS}:(1/2,1/2)$, $\Omega_{gde}=1$, $q=-1$, $\omega_{gde}=-1$. This point corresponds to a de Sitter solution since, $$\zeta=\frac{1}{2}\;\Rightarrow\;\dot H=0\;\Rightarrow\;H=H_0=const.,$$ while $\xi=1/2\;\Rightarrow\;H=H_0=\alpha/3$. It can be associated with a transient stage of the cosmic evolution since, as long as the eigenvalues of the linearization matrix $\lambda_1=2$, $\lambda_2=-3$, are of opposite sign, $P_{dS}$ is a saddle critical point.

\item Empty space critical point $P_E:(0,1),\;q=1$. The quantities $\Omega_{cdm}$, $\Omega_{gde}$, and $\omega_{gde}$ are undefined at $P_E$, however $\omega_{gde}\Omega_{gde}=0$. This equilibrium point corresponds to empty space static solution since $$\xi=0\;\Rightarrow\;H=0\;\Rightarrow\;\rho_{gde}+\rho_{cdm}=0,$$ and, besides, $H=0$ $\Rightarrow\;a(t)=a_0$. It is a future attractor since the (real) eigenvalues of the linearization matrix $\lambda_1=-1$, $\lambda_2=-2$, are both negative. As seen from figure \ref{fig1} this is not a global attractor but a local one: there is a non-empty set of orbits in $\Psi_\epsilon$ generated by appropriate initial data which lie above the separatrix $\sigma$ (\ref{separatrix}) and are attracted towards $P_E$, the remaining orbits go elsewhere. 

\end{enumerate}

\subsection{Remarkable non-equilibirum points in $\Psi_\epsilon$}

Besides the two critical points of (\ref{ode-e}) above there are two other points in the phase space $\Psi_\epsilon$ which are not equilibrium points but have interesting properties and deserve separate discussion.

\begin{enumerate}

\item Dark matter dominated non-equilibrium point $$P_{MD}:\left(1,\frac{4}{5}\right),\;\Omega_{cdm}=1,\;q=\frac{1}{2},$$ where $\omega_{gde}$ is undetermined. At this point the cosmic dynamics is dictated by the cosmological equations $$3H^2=\rho_{cdm},\;\dot H=-\frac{\rho_{cdm}}{2},\;\rho_{cdm}\propto a^{-3}.$$ 

Orbits in the phase space seem to emerge from $P_{MD}$ as it were a past attractor, however, this point is not even a critical one. Actually, at $(1,4/5)$, $$\xi'=0,\;\zeta'=\text{undefined}.$$ Orbits seem to start at $P_{MD}$ because the starting point of the cosmic evolution in general relativity is the big bang event. I. e., a state characterized, in particular, by $H\rightarrow\infty$, which means that $\xi\rightarrow 1$. Hence the starting point of physically meaningful orbits in the phase space should belong in ${\bf v}$. But, as it will be shown below, only the point $(1,4/5)$ can be associated with matter fields which are compatible with general relativity. Otherwise, for $(1,\zeta),\,\zeta\neq 4/5$ the ghost dark energy component behaves as 'pure pressure' (vanishing energy density).\footnote{If in the cosmological equations (\ref{cosmo-eq}) add a radiation component with energy density $\rho_r$ and barotropic pressure $p_r=\rho_r/3$: $3H^2=\rho_r+\rho_{cdm}+\rho_{gde}$, then the orbits would start at a non-equilibrium radiation dominated point. In that case the matter-dominated point would not be remarkable at all.} 

If consider small perturbations around $P_{MD}$: 

\bea &&\xi\rightarrow 1-\delta\xi,\;\zeta\rightarrow\frac{4}{5}+\delta\zeta\;\Rightarrow\nonumber\\
&&(\delta\xi)'\approx\frac{3}{2}\,\delta\xi\,\Rightarrow\,\delta\xi(\tau)=\delta\xi(0)\,e^{3\tau/2},\label{xi-pert}\\
&&(\delta\zeta)'\approx\frac{6}{25}-\frac{27}{5}\delta\zeta+\frac{4\delta\zeta+5\delta\zeta^2}{\delta\xi}.\nonumber\eea The solution of the latter differential equation is found to be:\footnote{In order to solve this equation it might be useful to perform the following change of parameter: $\tau\rightarrow 2\,e^{-3\tau/2}/3$.}

\bea &&\delta\zeta(\tau)=\frac{-100 C\,\text{U}(a,b,z)-4\text{M}(a,b,z)}{H(z)},\label{zeta-pert}\\
&&H(z)\equiv 125 C\,\text{U}(a,b,z)+475 C\,\text{U}(\bar{a},b,z)\nonumber\\
&&\;\;\;\;\;\;\;\;\;\;\;\;\;\;\;\;\;\;\;\;+5\text{M}(a,b,z)-95\text{M}(\bar{a},b,z),\nonumber\eea where $$z=\frac{8 e^{-3\tau/2}}{3\delta\xi(0)},\;a=-\frac{19}{5},\;\bar{a}=-\frac{14}{5},\;b=-\frac{13}{5},$$ $C$ is an arbitrary constant, and 

\bea &&\text{M}(a,b,z)=\frac{\Gamma(b)}{\Gamma(a)\Gamma(b-a)}\int_0^1 e^{zt} t^{a-1}(1-t)^{b-a-1}dt,\nonumber\\
&&\text{U}(a,b,z)=\frac{1}{\Gamma(a)}\int_0^\infty e^{-zt} t^{a-1}(1+t)^{b-a-1}dt,\nonumber\eea are the confluent hypergeometric functions (also known as Kummer's functions) of the 1rst and 2nd kind respectively.

In the figure \ref{fig2} we show the evolution of these perturbations for $\delta\xi(0)=0.1$ ($C=0.1$): $\delta\xi$ vs $\tau$ is depicted in the left-hand upper panel, while in the right-hand one $\delta\zeta$ vs $\tau$ is shown. Notice that, while the $\xi$-perturbation $\delta\xi$ exponentially increases with $\tau=\ln a$, the perturbation $\delta\zeta$ undergoes a dramatic almost sudden increase at a given value $\tau=\tau_0$ (in the figure $\tau\approx 1.2$). This means that the dynamical system very quickly departs from the non-equilibrium point $P_{MD}$. In other words, the non-equilibrium matter-dominated point is highly unstable.

\item Big rip non-equilibrium point $P_{BR}:(1,0)$. It is also remarkable since orbits in phase space which lie below the separatrix $\sigma$ (\ref{separatrix}) end up at $P_{BR}$. This is so not because the point is an attractor (it is not even an equilibrium point), but because at $P_{BR}$ the phase space itself shrinks to a point 'focusing' all of the orbits.

At this point $\Omega_{cdm}$, $\Omega_{gde}$, $\omega_{gde}$, and $q$ are all undefined quantities. In general, as one approaches to $P_{BR}$ (see Eq.(\ref{useful-1})) $$\frac{\dot H}{2H^2}\rightarrow\infty,\;H\rightarrow\infty,\;\omega_{gde}\rightarrow -\infty,\;q\rightarrow -\infty,$$ while the the Hubble horizon shrinks to a point $\tilde r_T\rightarrow 0$.

\end{enumerate}

The fact that $P_{MD}$ and $P_{BR}$ are non-equilibrium points means that: i) the phase space orbits do not really start at $P_{MD}$ but at some other point outside of the physical phase space $\Psi_\epsilon$, and ii) those orbits which lie below the separatrix do not really end up at $P_{BR}$ but they are continued into the unphysical region outside $\Psi_\epsilon$.

\subsection{Phase space orbits}

In the figure \ref{fig1} a set of phase space orbits in $\Psi_\epsilon$ is shown. These are separated into orbits which entirely lie above the separatrix $\sigma$ (\ref{separatrix}) and those which lie below $\sigma$. As one goes back into $\tau$-time these orbits are focused into the matter-dominated non-equilibrium point $P_{MD}$.

The orbits above the separatrix (dot-dashed curve in the figure \ref{fig1}), depending on the initial conditions, may either approach towards the local attractor $P_E$ -- which is associated with empty static universes -- into the future, or they may go elsewhere in the upper horizontal edge ${\bf h}$ which can be associated with ghost dark energy in the very peculiar form of 'pure pressure' ($\rho_{gde}=0$, $p_{gde}\neq 0$). This kind of source of gravity is not consistent with general relativity. Actually, suppose a FRW spacetime (flat spatial sections as before) is filled with such a 'pure pressure' thing. Then one would have

\bea 3H^2=0,\;2\dot H+3H^2=-p\;\Rightarrow\;H=0,\;\dot H\neq 0,\nonumber\eea which is a nonsense as long as the parametric pressure $p$ is non-vanishing. Hence, one should take into consideration only those orbits which end up at the local attractor $P_E$. Besides, only those orbits which hit the small region in the phase space bounded by: i) the horizontal line $\zeta=2/3$, ii) the oblique line $\zeta=1-\xi$, and iii) the separatrix $\sigma$: $\zeta=(7\xi-3)/(8\xi-3)$, are to be considered as (would be) adequate cosmological models depicting a stage of accelerated expansion (recall that the deceleration parameter $q<0$ only for $\zeta<2/3$). If one wants to be more precise, since the present stage of the cosmic expansion is quite well approximated by the $\Lambda$CDM model, then one should take into serious consideration only those orbits which approach close enough to the de Sitter saddle critical point $P_{dS}:(1/2,1/2)$.

Orbits which lie below the separatrix are inevitably focused into the future towards the non-equilibrium point $P_{BR}:(1,0)$ which is associated with a catastrophic big rip event. The most interesting feature of these orbits is that they do the crossing of the phantom divide $\omega_{gde}=-1$ at $\zeta=1/2$ (doted horizontal line in the figure \ref{fig1}). Using the same argument as in the case of orbits which lie above the separatrix, in the present case one should take into serious consideration only those orbits which approach close enough to the de Sitter saddle critical point $P_{dS}:(1/2,1/2)$. For those orbits, until the de Sitter point $P_{dS}$ is approached, the ghost dark energy component of the cosmic mixture approximately behaves as a cosmological constant since along the separatrix $\omega_{gde}=-1$.

An evident drawback of this cosmological model is the absence of an equilibrium point associated with matter dominance. As shown above, the non-equilibrium matter-dominated point $P_{MD}:(1,4/5)$ is a very unstable configuration since, according to equations (\ref{xi-pert}), (\ref{zeta-pert}), any small perturbation around $P_{MD}$ grows exponentially along the $\xi$-direction, while along the $\zeta$-direction it undergoes a sudden extraordinary increase at some $\tau=\tau_0$, which very quickly takes the autonomous system far apart from the matter-dominated point (see the figure \ref{fig2}). At this $\tau_0$ the 'velocity' component $\delta\zeta'(\tau_0)$ is much larger than $\delta\xi'(\tau_0)$, which means that the orbits through $P_{MD}$ are practically vertical curves in the figure \ref{fig1} which: i) do not approach to the de Sitter point under any circumstances, and ii) very quickly reach either points in the horizontal edge which are not compatible with general relativity, or approach to the big rip solution after an early crossing of the phantom divide $\omega_{gde}=-1$. This is illustrated in the figure \ref{fig3} where a set of orbits generated by initial data in the neighborhood of the matter-dominated non-equilibrium point $P_{MD}:(1,4/5)$ is shown.

A transient stage of matter dominance -- responsible for the observed amount of cosmic structure -- is an essential ingredient of the accepted cosmological paradigm. Hence, on the basis of the absence of a matter-dominated equilibrium point in $\Psi_\epsilon$ the QCD GDE cosmological model studied here may be ruled out as it is unable to explain the formation of structure in our universe.

\section{Discussion and Conclusion}

According to the accepted cosmological paradigm the main features of the cosmic history of our universe are: i) an early period of inflation, ii) density inhomogeneities produced from quantum fluctuations during inflation, iii) a flat, critical density, acceleratingly expanding universe, iv) energy density budget consisting of roughly $2/3$ of dark energy (DE) and $1/3$ of cold dark matter (CDM), and v) matter content: $29\%$ -- CDM, $4\%$ -- baryons, and $0.3\%$ -- neutrinos. Any successful cosmological model should be able to explain the above mentioned features. In particular a transient stage of matter-dominance is mandatory to explain the observed amount of cosmic structure. 

In this paper we have demonstrated that, as a matter of fact, the so called Veneziano (also QCD) ghost dark energy model \cite{urban,ariel,ohta,chinos,chinos',also,instability,alberto-rozas,other,gde-thermodynamics,iterms-also,global-behaviour,low-e-qcd,gde-motivation,gde-motivation'} can not be a successful cosmological model in the sense mentioned above, since there is not any critical point in the equivalent phase space of the model which can be associated with a transient stage of matter-dominance. There are only two critical points: i) the saddle equilibrium point corresponding to (accelerated) de Sitter expansion where the ghost dark energy behaves as a cosmological constant, and ii) the empty static universe which is a local attractor in the phase space. This does not mean that a particular solution of the cosmological equations where matter dominates could not be picked out under specific initial conditions. What this really means is that in case such a particular solution is found it will be very unstable, so that it would not last for enough time as to account for the observed amount of cosmic structure. In the present case this is easily illustrated by considering small perturbations around $P_{MD}$: $(1,4/5)\,\rightarrow\,(1-\delta\xi,4/5+\delta\zeta)$. 

As shown in section \ref{time-d} (see the figure \ref{fig2}) the evolution of the perturbations can be summarized as it follows. The component $\delta\xi$ grows exponentially ($\delta\xi(\tau)\propto e^{3\tau/2}$), while the component $\delta\zeta$ undergoes a sudden uncontrollable huge increase at some time $\tau_0$. This means that orbits originated from data in the vicinity of $P_{MD}$ will very quickly reach either points in the upper horizontal edge ${\bf h}$ which are associated with 'pure pressure' GDE sources not compatible with general relativity, or will do an early crossing of the phantom divide to approach to the big rip singularity (see the figure \ref{fig3}). In either case the resulting orbits will not approach to the de Sitter saddle point. The associated cosmological evolution will appreciably depart from that predicted by the $\Lambda$CDM model at any stage. 

Non-equilibrium points are meet only under very specific and unique arrangement of the initial conditions. This is to be contrasted with the fact that for critical points there can be a non-empty (perhaps infinite) set of initial conditions leading to orbits which meet the (vicinity of the) point: i) for past/future attractors any set of initial conditions picks a congruence of orbits emerging from the past attractor/focusing into the future attractor, and ii) for a saddle equilibrium point there is also a non-empty set of initial conditions leading to orbits which approach close enough to it. A qualitative analysis can rely on the 'speed' $v=\sqrt{x'^2+y'^2+...}$ at which a given orbit approach to a phase space point as a parameter to judge on the $\tau$-interval, $\Delta\tau\propto v^{-1}$, the orbit spends in the vicinity of the point. In the vicinity of critical points, since $x'=y'=...\approx 0$, the speed $v_{crit}$ of the orbit is vanishingly small, and $\Delta\tau_{crit}$ is large. For ordinary (non-equilibrium) points since $x'\neq 0,\;y'\neq 0,...,$ etc., the speed $v_{ord}$ is a non-vanishing quantity and, hence $$\frac{\Delta\tau_{ord}}{\Delta\tau_{crit}}=\frac{v_{crit}}{v_{ord}}\ll 1.$$ 

If associate a given orbit with a pattern of cosmological evolution, then ($\Delta\tau=\Delta a/a$) $$\frac{\Delta\tau_{ord}}{\Delta\tau_{crit}}=\frac{\Delta z_{ord}}{\Delta z_{crit}}\ll 1,$$ where $z$ is the red shift. One can safely say that the Universe spends much more amount of redshift in a stage described by the solution associated with a critical point than in a stage described by the solution associated with a non-equilibrium point as it is the matter-dominated point $P_{MD}$ in the present case. This is the most one can say based on the results of the dynamical systems study.

From the results of the study in section \ref{propto-h} it can be inferred that the simplified QCD GDE model where $\rho_{gde}=\alpha H$ is suitable from the cosmological point of view since, besides the de Sitter point (future attractor) there are critical points associated with radiation-dominance (past attractor) and cold dark matter-dominance (saddle point). Notice, however, that there is no way in which the latter points can be retrieved from the more realistic model where $\rho_{gde}=\alpha H(1-\epsilon)$ in the limit when $\epsilon\rightarrow 0$. This does not mean that our study is not mathematically consistent (it is), but rather that one can not pretend to describe the entire cosmic history with an approximate model which is valid only during a particular stage of GDE dominance ($\epsilon=0$). Hence the radiation-dominated and CDM-dominated critical points in section \ref{propto-h} do not actually belong in the phase space of the Veneziano ghost dark energy model.

We conclude that there are at least two good reasons why we should rule out the Veneziano ghost as a model for the dark energy: i) it is plagued by classical instability against small perturbations of the background due to negativity of the sound speed squared, and ii) it is unable to explain the formation of structure in our universe. While the former drawback above might be circumvented through more or less plausible arguments \cite{alberto-rozas,ariel,global-behaviour} the latter one is conclusive. 

The authors thank SNI of Mexico for support. The work of R G-S was partly supported by SIP20120991, SIP20131811, COFAA-IPN, and EDI-IPN grants. I Q and M T-M thank "Programa PRO-SNI, Universidad de Guadalajara" for support under grant No 146912.

\end{document}